\documentclass[aps,pra,superscriptaddress,showpacs, twocolumn,10pt]{revtex4-1}

\usepackage{amsfonts,amssymb,amsmath}
\usepackage{graphics,graphicx,epsfig}
\usepackage{amsthm}
\usepackage{epstopdf}
\usepackage{color}

\newcommand{\ket}[1]{ | #1 \rangle}

\newcommand{\Tr}{\mathrm{Tr}}

\renewcommand{\epsilon}{\varepsilon}

\begin{document}

\title{Correlation-based entanglement criterion in bipartite multiboson systems}

\author{Wies\l aw~Laskowski}
\affiliation{Institute of Theoretical Physics and Astrophysics, University of Gda\'nsk, PL-80-952 Gda\'nsk, Poland}

\author{Marcin~Markiewicz}
\affiliation{Faculty of Physics, University of Warsaw,  ul. Pasteura 5, PL-02-093 Warszawa, Poland}
\affiliation{Institute of Theoretical Physics and Astrophysics, University of Gda\'nsk, PL-80-952 Gda\'nsk, Poland}

\author{Danny~Rosseau}
\affiliation{National Institute of Informatics, 2-1-2 Hitotsubashi, Chiyoda-ku, Tokyo 101-8430, Japan}

\author{Tim~Byrnes}
\affiliation{New York University, 1555 Century Ave, Pudong, Shanghai 200122, China}
\affiliation{NYU-ECNU Institute of Physics at NYU Shanghai, 3663 Zhongshan Road North, Shanghai 200062, China}
\affiliation{National Institute of Informatics, 2-1-2 Hitotsubashi, Chiyoda-ku, Tokyo 101-8430, Japan}

\author{Kamil~Kostrzewa}
\affiliation{Institute of Theoretical Physics and Astrophysics, University of Gda\'nsk, PL-80-952 Gda\'nsk, Poland}

\author{Adrian~Ko{\l}odziejski}
\affiliation{Institute of Theoretical Physics and Astrophysics, University of Gda\'nsk, PL-80-952 Gda\'nsk, Poland}

\begin{abstract}
We describe a criterion for the detection of entanglement between two multi-boson systems. The criterion is based on calculating correlations of Gell-Mann matrices with a fixed boson number on each subsystem.  This applies naturally to systems such as two entangled  spinor Bose-Einstein condensates.  We apply our criterion to several experimentally motivated examples, such as an $ S^z S^z $  entangled BECs, ac Stark shift induced two-mode squeezed BECs, and photons under parametric down conversion. We find that entanglement can be detected for all parameter regions for the most general criterion.  Alternative criteria based on a similar formalism are also discussed together with their merits. 
\end{abstract}

\pacs{03.67.Mn}

\maketitle

\section{Introduction}

The notion of quantum entanglement is closely connected to the tensor product structure of a state space of composite quantum systems. For some types of composite systems that lack this structure, entanglement is no longer uniquely defined. This happens for systems of indistinguishable particles, in which symmetrization or anti-symmetrization of the wave function imposes some sort of ``default'' entanglement. As an elementary example of this, consider two identical bosons which occupy two quantum states, for example atomic levels denoted $ a $ and $ b $ of an atom.  As they are indistinguishable bosons, it is appropriate to use the second quantized notation to write the quantum state as 
\begin{align}
a^\dagger b^\dagger | 0 \rangle , 
\end{align}
where $ a,b $ are bosonic annihilation operators of the states. This does not have the appearance of possessing any entanglement.  In first quantized notation, one would however write
\begin{align}
\frac{1}{\sqrt{2}} \left( \psi_a (r_1) \psi_b (r_2) + \psi_a (r_2) \psi_b (r_1) \right),
\end{align}
which has the appearance of an entangled state.  The amount of entanglement in such systems that is operationally accessible is difficult to define and has been a subject of a long debate. Initially the debate was motivated by the problem of nonclassical properties of a two-mode single photon field \cite{TWC91, Hardy94, GHZ95, Enk05}. Later many propositions for defining entanglement of indistinguishable particles were proposed \cite{PY01, SCKLL01, LZLL01, GF02, Shi03, Marzolino10, Marzolino11, Marzolino12, Marzolino14}, which were concluded by showing, that only a part of entanglement coming from symmetrization can be operationally extracted \cite{WV03, GM04, DDW06, Toth12, Toth14}. Recently more efficient way of extracting this kind of entanglement was proposed by a method called \emph{mode splitting} \cite{KCP14}. 

Understanding entanglement in indistinguishable systems is important to understand not only from a fundamental perspective, but increasingly in the context of technological applications.  The resource capabilities in quantum optical systems has been a topic of long-standing investigation, due to their highly coherent properties making them a natural candidate for quantum information tasks \cite{WV03}. 
Recent experimental advances in coherent control of Bose-Einstein condensates (BECs) have opened the door for utilizing these systems for tasks such as quantum metrology \cite{gross12}, interferometry \cite{muntinga14,chiow11}, quantum simulation \cite{bloch12}, and quantum computing \cite{byrnes15}.  In particular, the ability of realizing BECs on atom chips allows for the possibility of excellent coherent control \cite{bohi09}, and performing tasks such as spin squeezing \cite{riedel10}.  This allows for a naturally scalable architecture where multiple BECs could be produced on the same chip, allowing for the possibility of entanglement between BECs \cite{pyrkov13,hussain14,rosseau14,abdelrahman14}.  While entanglement between two BEC has not been realized to date, entanglement between a BEC and an atom has been demonstrated \cite{lettner11}. While entanglement criteria for optical systems has been relatively well-studied \cite{duan00,walborn09}, the analogous case for BECs has not been performed to the same level \cite{byrnes13}.  

In this paper we consider a hybrid scenario, in which two systems of bosons constitute two physically distinguishable components. For simplicity we assume that both components contain the same fixed number $N$ of indistinguishable particles, which can be prepared in two different modes. In such a  physical system, bipartite entanglement between the two components is well defined, and can be experimentally detected and used as a resource. The main goal of our paper is to present correlation-based criteria for detection of this kind of entanglement. A correlation based approach \cite{BBLPZ08, LMPZ11, MLPZ13, LMPW13} is desirable from an experimental point of view as they are most closely related to measured observables. However it cannot be directly applied to the case of multi-boson states, because it involves measurements performed on single localized particles. In order to extend applicability of correlation-based entanglement detetection to multiboson systems, we define the measurement basis for the two components, which enables us to treat the entire system as two effective $(N$+$1)$-level systems. Having defined the basis we apply quadratic correlation tensor criterion from \cite{BBLPZ08}, which allows for the efficient detection of bipartite entanglement between the two components. Note that the above method of probing entanglement utilizes the entire structure of subsystems that can be defined within the components. Furthermore, we demonstrate that using this entire structure is a necessary condition for detecting entanglement -- the compound observables with higher-dimensional eigenspaces (e.g. total angular momentum observables) lead to weaker and unsatisfactory conditions. Despite its simplicity, the model presented in this paper can be used to characterize entanglement in two experimentally significant physical systems: entangled spinor Bose-Einstein condensates \cite{B13} and non-classical light created during spontaneous parametric down conversion (SPDC) \cite{PCZWZ08}. We choose our examples based on realistic methods that could be used to generate entanglement in BECs and optical systems, which provide a test-bed for examining macroscopic entanglement \cite{V08, B13}.

This paper is structured as follows.  In Sec. \ref{sec:entcond} we introduce our approach for bipartite bosonic entanglement.  In Sec. \ref{sec:examples} we give various examples of our scheme applied to specific states, including the maximally entangled state, parametrically down converted state, two BECs evolving under a $ S^z S^z $ interaction, and two BECs in a two-mode squeezed state.  In Sec. \ref{sec:comp} we contrast our approach to alternative methods based on spin operators.  
We finally show our conclusions in Sec. \ref{sec:conc}.

\section{Entanglement condition}
\label{sec:entcond}

We first recall the simplest version of the geometric entanglement condition  \cite{BBLPZ08} in the context of two $d$-dimensional systems. 
Arbitrary quantum state of the entire system can be uniquely described by its correlation tensor $T$, which  is defined in the Gell-Mann matrix basis in a natural way \cite{HE81}
\begin{eqnarray}
\rho &=& \frac{d-1}{2d} ( M_0 \otimes M_0  + \sum_{i,j=1}^{d^2-1} T_{ij} M_i \otimes M_j \\
     &+& (d-1)\sum_{i=1}^{d^2-1} (T_{0i} M_{0} \otimes M_i + T_{i0} M_{i} \otimes M_0)), \nonumber
\end{eqnarray}
where the coefficients are
\begin{equation}
T_{ij} = \frac{d}{2(d-1)} {\Tr}(\rho M_i \otimes M_j),
\label{ctensor}
\end{equation}
where $M_i$ $(i=0,\dots, d^2-1)$ are the generalized Gell-Mann matrices (see e.g. \cite{BERTLMANN}) and $M_0 = \frac{1}{\sqrt{d(d-1)/2}} \openone$. The Gell-Mann
matrices generalize the previous approach considered in Refs. \cite{BBLPZ08, LMPZ11, MLPZ13, LMPW13} which were more focused on the $ \mbox{SU}(2)$ case to $ \mbox{SU}(d)$.  

The correlation tensors are elements of a real vector space with a usual scalar product
\begin{equation}
(X,Y)=\sum_{i,j} X_{ij} Y_{ij}.
\label{scalar}
\end{equation}
A state $\rho$ is separable if it can be expressed as a convex sum of product states
\begin{equation}
\rho_{\rm sep} = \sum_{i} p_i \rho^i_A \otimes \rho_B^i,
\end{equation}
with $p_i \geq 0$ and $\sum_i p_i =1$. In the language of the correlation tensors this decomposition reads $T^{\rm sep}  = \sum_i p_i T_i^{\rm prod}$, where $T_i^{\rm prod}  = T_i^{(A)} \otimes T_i^{(B)}$ and $T_i^{(k)}$ represents a single qu$d$it state.  The separable correlation tensors form a convex set. In a consequence, when $\rho$ is endowed by $T$, one has the following implication:
\begin{equation}
\rho~{\rm is~separable} \Rightarrow \exists_{T^{\rm prod}} (T, T^{\rm prod}) \geq (T,T),
\end{equation}
or, equivalently,
\begin{equation}
\max_{T^{\rm prod}} (T, T^{\rm prod}) < (T,T) \Rightarrow \rho~{\rm is~entangled}. 
\label{gen_crit}
\end{equation}
The proof of the above implications is presented in \cite{BBLPZ08}. The maximization of the left-hand side of (\ref{gen_crit}) is given by the highest Schmidt coefficient, $T_{\rm max}$, of the tensor $T$ \cite{footnote1}. Therefore, the condition
\begin{equation}
\epsilon = \frac{\mathcal{T}}{T_{\rm max}} >1 \Rightarrow \rho~{\rm is~entangled},
\label{crit_tmax}
\end{equation}
where $\mathcal{T} = (T,T) = \sum_{i,j}T_{ij}^2$, is a simple entanglement witness. Note that the determination of $T_{\rm max}$ requires knowledge of the entire tensor $T$ and in general can be a hard computational problem. Since $T_{\rm max}\leq 1$, we can rewrite the condition (\ref{crit_tmax}) as
\begin{equation}
\mathcal{T} >1 \Rightarrow \rho~{\rm is~entangled}.
\label{crit_1}
\end{equation}
Note that the original form of the entanglement criterion presented above is formulated only for the situation in which a subsystem is associated with a single $d$-level particle, and cannot be directly applied to the case subsystems consist of many indistinguishable particles.

Now we adapt the entanglement criterion to the case of bosonic systems.  Let us consider the most general state of a two-component system of $N$ two-mode bosons
\begin{equation}
\ket{\psi}= \sum_{k=0}^N  \sum_{l=0}^N  c_{kl} \ket{N-k}^a_{A}\ket{k}^b_{A}\otimes \ket{N-l}^a_{B}\ket{l}^b_{B},
\label{state_gen}
\end{equation}
where $\ket{i}^{\alpha}_{X}$ denotes the $i$-particle state in the spatial mode $X$ and some bosonic mode $\alpha$. The states within given spatial modes are given in particle number representation, whereas the tensor product symbol between the spatial modes indicates full distinguishability between them. For simplicity, we will use the abbreviated notation $\ket{i}^x_{X} \ket{j}^y_{X} \equiv \ket{i,j}_X$ and omit the tensor product symbol.
The core of our construction is the following definition of a new basis:
\begin{equation}
|k) \equiv \ket{N-k,k},
\label{newbasis}
\end{equation}
where the vector $|k)$ is defined as the $(k+1)$-th eigenvector of the last (diagonal) generalized  Gell-Mann matrix of dimension $N+1$. The definition (\ref{newbasis}) maps the two mode system with fixed number $N$ of particles into a $(N+1)$-level system, which can be understood as a single logical qu$d$it, where $d=N+1$.
In the new basis the state \eqref{state_gen} takes the form:
\begin{equation}
|\psi\rangle= \sum_{k=0}^N \sum_{l=0}^N  c_{kl} |k)_{A} |l)_{B},
\label{psi_gen_newbasis}
\end{equation}
and is formally equivalent to a state of two qu$d$its, with $d=N$+$1$.
As entanglement criterion we take the conditions (\ref{crit_tmax}) or (\ref{crit_1}) with the summation indices $i,j = \{1, \dots, N^2 + 2N \}$.

A specific example of (\ref{psi_gen_newbasis}) is the maximally entangled state of two qu$d$its (with $d=N+1 \geq 2$), which can be written 
\begin{equation}
|\psi^{(N)}_{\rm max})= \frac{1}{\sqrt{N+1}}\sum_{k=0}^N  |k)_{A} |k)_{B}.
\label{statemax}
\end{equation}
The entanglement identifier for this state can be evaluated to be 
\begin{equation}
\mathcal{T}^{(N)} = N+2
\end{equation}
and $T_{\rm max}=1$ \cite{LMPW13}. Since $\epsilon^{(N)} = N+2>1$, we see that it is entangled for all $N>0$, as expected. 

\section{Example applications}
\label{sec:examples}

Here we present a few examples of applications of the condition (\ref{gen_crit}) to detect entanglement of various classes of states. Our choices of examples are motivated by experimental realizability.  We show one example related to photonic systems showing parametric down conversion, and two examples for the BEC case involving an $ S^z S^z $ entangling gate and two-mode squeezing induced by entanglement with coherent light.

\subsection{Two BECs entangled with an $ S^z S^z $ interaction}

For our first example, we examine the case where two BECs are entangled using an $ S^z S^z $ interaction.  Several proposals for such an interaction exist, including photon mediated cavity based methods \cite{pyrkov13,hussain14,abdelrahman14} and interactions with state-dependent forces \cite{treutlein06}.  The type of entanglement that is produced by such an interaction has an exotic structure with fractal properties (the ``devil's crevasse'') \cite{byrnes13,kurkjian13}.  

To produce the state, two spin coherent states are initially polarized in the $ S^x $ direction, and have a $ S^z S^z $ interaction applied to them.  For BECs a general spin coherent state is written \cite{B13}
\begin{eqnarray}
|\alpha, \beta \rangle \rangle &=& \frac{1}{\sqrt{N!}} (\alpha a^\dagger+ \beta b^\dagger)^N |0,0\rangle \\
&=& \sum_{k=0}^N \sqrt{{N \choose k}} \alpha^{N-k} \beta^k |N-k,k \rangle.
\end{eqnarray}
We define the Schwinger boson operators as 
\begin{align}
S^x & = a^\dagger b +b^\dagger a \nonumber \\
S^y & = -i a^\dagger b + i b^\dagger a \nonumber \\
S^z & = a^\dagger a - b^\dagger b.  \label{schwingerboson}
\end{align}
After the $S_1^z S_2^z$ entangling operation we obtain \cite{B13}:
\begin{eqnarray}
&&\ket{\psi_{\rm BEC}} = e^{-i S_1^z S_2^z \tau}  \Big|\frac{1}{\sqrt{2}},\frac{1}{\sqrt{2}} \Big\rangle \Big\rangle \Big|\frac{1}{\sqrt{2}},\frac{1}{\sqrt{2}} \Big\rangle \Big\rangle \nonumber \\
&&= \frac{1}{\sqrt{2^N}} \sum_{k=0}^N \sqrt{{N \choose k}} \Big | \frac{e^{i (N-2k)\tau}}{\sqrt{2}}, \frac{e^{-i (N-2k)\tau}}{\sqrt{2}}\Big\rangle \Big \rangle |N-k,k\rangle \nonumber \\
&&=  \sum_{k,l=0}^N b_{kl}  |N-k,k\rangle |N-l,l\rangle, \label{BECstate}
\end{eqnarray}
where $b_{kl}=\frac{1}{2^N}\sqrt{{N \choose k} {N \choose l}} e^{-i (N-2k)(N-2l)\tau}$.  The interaction is periodic in time $ \tau = \pi/2 $ hence we consider only $\tau = [0,\pi/2)$.

When the number of particles per BEC is $ N = 1 $, the system is exactly equivalent to two qubits.  In this case, the non-vanishing elements of the correlation tensor are equal to $
T_{11}^{(1)} =1 \nonumber $ and $ T_{23}^{(1)} =T_{32}^{(1)}= \sin 2 \tau $.  We may then evaluate:
\begin{align}
\mathcal{T}^{(1)} = 1 + 2 \sin^2 2 \tau
\end{align}
and 
\begin{align}
T_{\rm max}^{(1)} = 1 . 
\end{align}
We therefore find that  $\epsilon^{(1)}>1$ for all $\tau \neq 0$. The maximal value of $\epsilon^{(1)} = 3$ is achieved for $\tau=\pi/4$, at which point we obtain a Bell state. All the expected behavior for an entanglement detector in this case is seen.  

\begin{figure}
\includegraphics[width=0.46\textwidth]{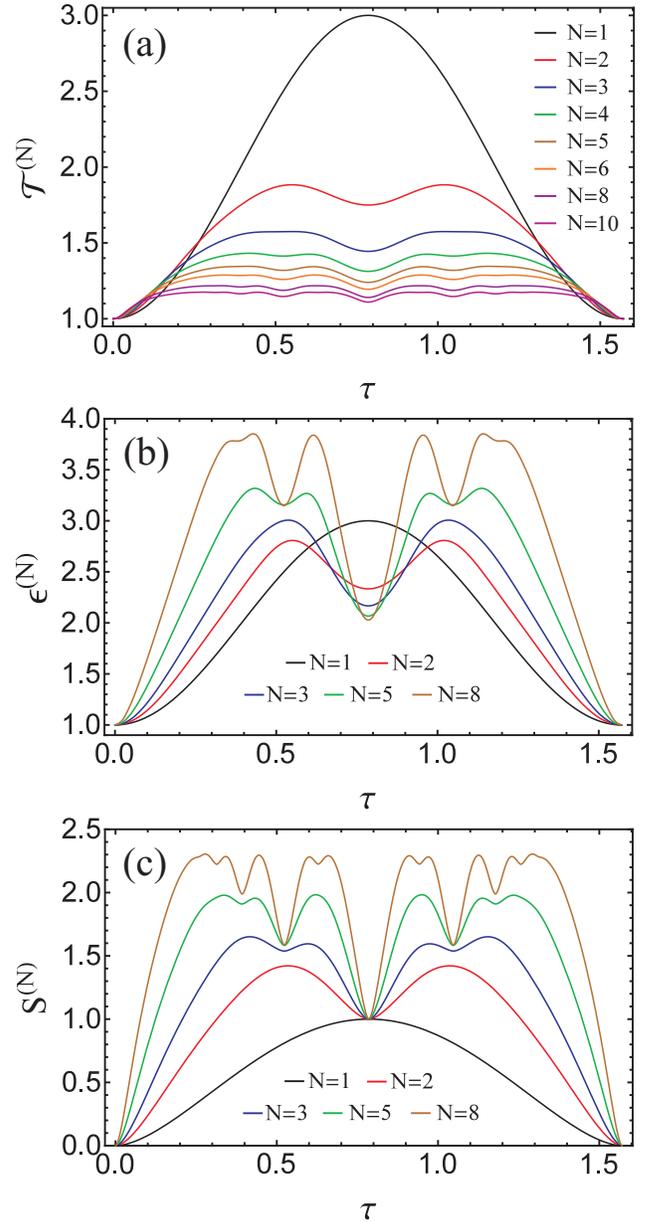}
\caption{\label{figT1} (Color
  online). Various entanglement identifiers (a) $\mathcal{T}^{(N)}$ (b) $\epsilon^{(N)}$ (c) the von Neumann entropy $ S $ for two BECs entangled via a $ S^z S^z $ interaction as a function of $\tau$.}
\end{figure}

For larger $ N $, the complexity of the entanglement detector increases.  For example, for $N=2$, we obtain 
\begin{align}
\mathcal{T}^{(2)}= 2 -\frac{c}{128}
\end{align}
and 
\begin{align}
T_{\rm max}^{(2)} = \frac{1}{2} + \frac{1}{16}\sqrt{c/2},
\end{align}
where $c = 53 + 48 \cos 4 \tau + 24 \cos 8 \tau + 3 \cos 16 \tau$.	We therefore illustrate our results numerically for $N \leq 8$, as shown in Fig. \ref{figT1}. In Fig. \ref{figT1}(a) we plot $\mathcal{T}^{(N)}$, which exceeds unity for all $\tau \neq 0$ (see Fig. \ref{figT1}).  While this shows that entanglement is present for all times by virtue of (\ref{crit_1}), the amount that this exceeds unity decreases with $ N $.  In fact the amount of entanglement, as measurement by the von Neumann entropy
\begin{align}
S = - \mbox{Tr} \left( \rho_1 \log_2 \rho_1  \right)
\end{align}
where $ \rho_1 $ is the density matrix with the partial trace over BEC 2, is known to increase with $ N $.  This is natural as the Hilbert space of the system increases with $ N $, hence the amount of entanglement can be also expected to grow.  The maximally entangled state (\ref{statemax}) has an entanglement equal to $ S_{\rm max} = \log_2 (N+1) $.  

To remedy this behavior one may instead look at the stricter criterion given according to (\ref{crit_tmax}).  Fig. \ref{figT1}(b) shows the behavior of $\epsilon^{(N)}$ with the entangling time.  We see that again this exceeds unity for all times indicating the presence of entanglement. In contrast to $\mathcal{T}^{(N)}$, we see that the amount the bound is exceeded by grows with  $N $, and more resembles the behavior of the von Neumann entropy. We may therefore observe that $\epsilon^{(N)}$ is a more sensitive indicator of entanglement, although it has the additional overhead of the necessity of calculating $ T_{\rm max} $. 

We note that there is a very simple criterion that can be used to detect entanglement, restricted to the pure states and any particle number $ N $.  The observation is that: {\em a pure state $\ket{\psi}$ is entangled if and only if the lengths of its local Bloch vectors are less than 1} \cite{NIELSEN}.
For the BEC state, two Bloch vectors are of the same length given by:
\begin{equation}
||\vec a^{(N)}||^2 = b^{N}_0  + \sum_{l=1}^n b^{N,l} \cos^{2n} 2 l \tau,
\end{equation}
where
\begin{eqnarray}
b^{N,l} &=&  \frac{2{\rm Cat}(N) (N+1) \prod_{i=2}^l (N-i+1)}{4^N \prod_{i=2}^l (N+i) }, \\
b^{N}_0 &=& \frac{(N+1)^2 {\rm Cat(N)} }{4^N N}-\frac{1 }{N}
\end{eqnarray}
and {\rm Cat} is the Catalan number.
For the first few values of $N$, $||\vec a^{(N)}||^2$ is equal to:
\begin{eqnarray}
||\vec a^{(1)}||^2&=&\cos^2 2 \tau\\
||\vec a^{(2)}||^2&=&\frac{1}{16} + \frac{3}{4} \cos^4 2\tau + \frac{3}{16} \cos^4 4\tau \nonumber\\
||\vec a^{(3)}||^2&=& \frac{1}{12} + \frac{5}{8} \cos^6 2 \tau + \frac{1}{4 }\cos^6 4 \tau + \frac{1}{24} \cos^6 6 \tau \nonumber\\
||\vec a^{(4)}||^2&=&\frac{47}{512} + \frac{35}{64} \cos^8 2 \tau + \frac{35}{128}  \cos^8 4 \tau \nonumber \\
&+& \frac{5}{64} \cos^8 6 \tau + \frac{5}{512} \cos^8 8 \tau \nonumber
\end{eqnarray}
It is easy to see that the length of the Bloch vectors $||\vec a^{(N)}||^2 = 1$ only for $\tau=\{0,\pi/2\}$. For other values of $\tau$, $||\vec a^{(N)}||^2 < 1$, which proves the entanglement of the BEC state. 

For $N=1$, there exists such $\tau = \pi/4$ that the state is maximally entangled $(||\vec a^{(1)}||^2 = 0)$. For $N=3,4$ the minimal value of $||\vec a^{(N)}||^2$ increases. For $N>4$ the length of the Bloch vectors is a decreasing function of $N$ and for $N \to \infty$, $||\vec a^{(N)}||^2 \to 0$, which means that the BEC state again becomes maximally entangled (in this limit).

\subsection{Two BECs entangled via two mode squeezing}

\label{nf}

We next consider an entanglement generating procedure where we adapt experimental techniques used to entangle two atomic ensembles to BECs. Specifically, we consider the protocol experimentally realized by Julsgaard {\it et al.} \cite{Julsgaard} and proposed in Refs. \cite{Kuzmich,duan00b}.  In this protocol, one starts with two BECs, both polarized in the $ S^x $ direction, which are illuminated by coherent light in a superposition of two polarization modes
\begin{align}
 | \frac{1}{\sqrt{2}}, \frac{1}{\sqrt{2}} \rangle \rangle  | \frac{1}{\sqrt{2}}, \frac{1}{\sqrt{2}} \rangle \rangle  | \alpha \rangle 
\end{align}
where the coherent state of light of amplitude $ \alpha $ is 
\begin{align}
| \alpha \rangle =  e^{-|\alpha|^2/2} e^{\alpha /\sqrt{2} (c^{\dagger} + d^{\dagger})} |0\rangle .  
\end{align}
and $ c,d $ denote the two polarization modes. The light and ensembles then interact via the ac Stark shift, as the light is detuned off the excited state resonance.   The interaction Hamiltonian is 
\begin{align}
H_{\rm ac} = (S_1^z + S_2^z) J^z
\label{hamac}
\end{align}
where $ S_{1,2}^z $ are the total spin operators as before, and $ J^z = c^\dagger c - d^\dagger d $ is the spin operator for the light.  This couples the three systems and produces entanglement of the type BEC-light-BEC.  One then measures the light in the $ S^y $ basis.  This yields the state \cite{Petterson}
\begin{eqnarray}
\ket{t, n_c, n_d} \propto \sum_{k_1,k_2} A_{n_c, n_d}(k_1,k_2) \ket{k_1,k_2},
\label{ncndstate}
\end{eqnarray}
where, up to an irrelevant phase, 
\begin{equation}
A_{n_c, n_d}(k_1,k_2) = \frac{1}{2^N} \sqrt{{N\choose k_1}{N\choose k_2}}  (-1)^{k_2} \sin^{n_c} \xi  \cos^{n_d} \xi,
\label{Afunc}
\end{equation}
and $\xi = {2t(k_1+k_2-N)+\frac{\pi}{4}}$. The proportionality on the state (\ref{ncndstate}) refers to the fact that the state is unnormalized as it is the state after a projective measurement has been made on the photons.  The $ n_{c,d} $ are the random measurement outcomes for the number of photons detected in the $ S^y $ basis.  

\begin{figure}
\includegraphics[width=0.48\textwidth]{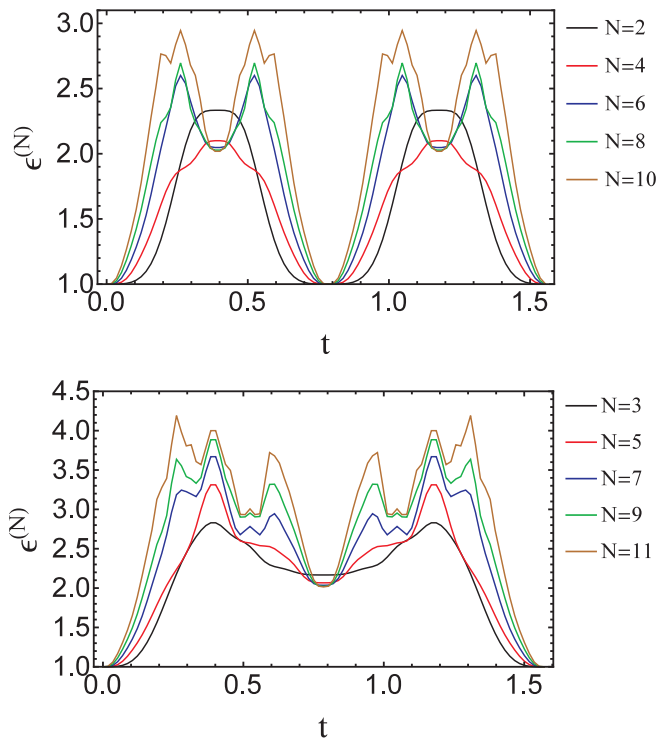}
\caption{\label{f_macro}(Color online). The values of $\epsilon^{(N)}$ for the macroscopic entanglement state (\ref{ncndstate}) as a function of $t$ for $N$ odd and even, separately.}
\end{figure}

We have performed numerical calculations for $N$ up to 40 atoms.  We assume parameters such that $ |\alpha|^2 = N $, which is the typical experimental regime \cite{Julsgaard}.  As the photon number in a coherent state should obey $ n_c + n_d \approx |\alpha|^2 $, we take $n_c = \left \lfloor \frac{N}{2} \right \rfloor$, $n_d= N - n_c$.  Fig. \ref{f_macro} shows how the entanglement parameter $\epsilon^{(N)}$ depends on the interaction time $t$ for various $N$. We find that there is a strong odd/even effect where the periodicity is $ \pi/4 $ for even $ N $ and $ \pi/2 $ for odd $ N $. This follows from the exponents in (\ref{Afunc}) where for even $ N $ our choice gives 
$ n_c = n_d $, and we have the simplification
\begin{align}
A_{n_c, n_d}(k_1,k_2) = & \frac{1}{2^{N+n_c}} \sqrt{{N\choose k_1}{N\choose k_2}}  (-1)^{k_2} \nonumber \\
& \times \cos^{n_c} \left(4t(k_1+k_2-N) \right) ,
\label{simpleA}
\end{align}
which has an extra factor of two in the cosine. For odd $ N $, we have an additional factor of $ \cos \xi $ in (\ref{simpleA}) which then has twice the periodicity of the even case.  In either case we find that our approach successfully detects entanglement, with $\epsilon^{(N)} > 1 $ for all $ t $.  

\begin{figure}
\includegraphics[width=0.44\textwidth]{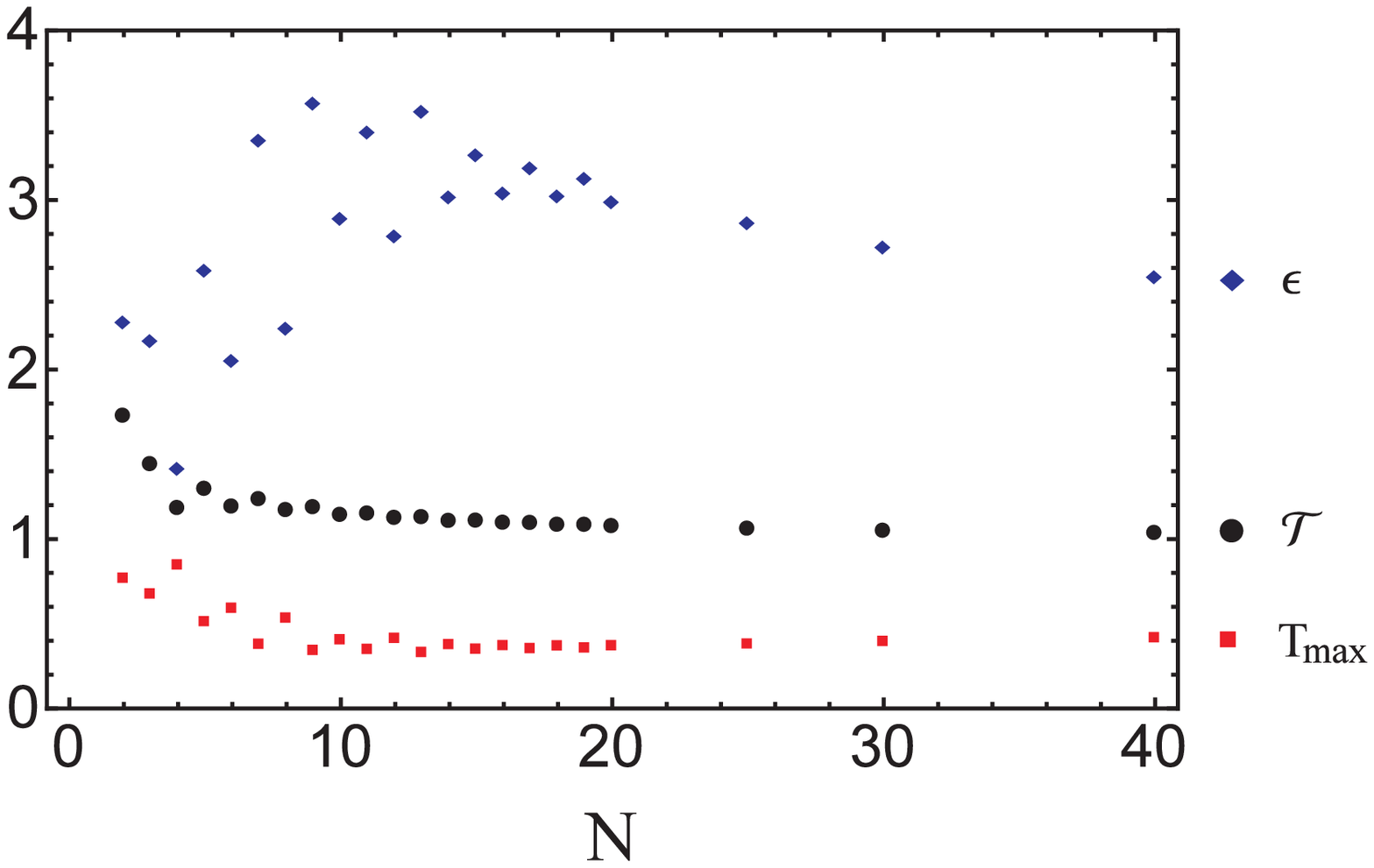}
\caption{\label{f_macro1}(Color online). The values of $\mathcal{T}^{(N)}$, $T_{\rm max}$ and $\epsilon^{(N)}$ for the macroscopic entanglement state (\ref{ncndstate}) with $t = 2.5/N$. For each case, $\epsilon^{(N)} > 1$ what proves entanglement.}
\end{figure}

In the experimental system, in fact the typical timescale of entanglement is the short-time limit, with typical values $t \sim 2.5 / N $ \cite{Julsgaard,Petterson}.  We plot $\epsilon^{(N)}$ in Fig. \ref{f_macro1}, as a function of $ N $.  We see that the amount that the bound is exceeded does not generally increase with $ N $, and it stays mostly in the vicinity of $\epsilon^{(N)} \sim 2$.  This is expected as in the short-time limit with interaction times in the region of $ t \sim 1/N $ the amount of entanglement is generally of the order of $ N = 1 $ case and remains a constant.  This behavior was observed also for the $ S^z S^z $ interactions \cite{byrnes13} where for interaction times of $ \tau \sim 1/N $ the von Neumann entropy is of order unity. This originates from the dimensions of the ac Stark shift Hamiltonian (\ref{hamac}) which is of order $ O(H) \sim N^2 $, and therefore to keep the total evolution $ e^{-iHt} $ equivalent the time needs to be scaled as $ t \sim 1/ N $.


\subsection{State from parametric down conversion}

\label{par}

Our final example is the resulting multiphoton state in the spontaneous parametric down conversion process creating an
entangled bright squeezed vacuum state \cite{PDCref}.  The state is
\begin{equation}
\ket{\psi_{\rm PDC}} = \frac{1}{\cosh^2 K} \sum_{N=0}^{\infty} \tanh^N K  \sqrt{N+1} ~\ket{\psi^{(N)}_{\rm max}},
\label{pdcstate}
\end{equation}
where
\begin{equation}
\ket{\psi^{(N)}_{\rm max}}= \frac{1}{\sqrt{N+1}}  \sum_{m=0}^N \ket{N-m,m}_{A}\ket{N-m,m}_{B},
\end{equation}
and $\ket{N-m,m}_{X}$ denotes a state of $N$ photons ($N-m$ horizontally and $m$ vertically polarized photons) in the spatial mode $X$ ($X=\{A,B\}$). 

As the number of photons in the state (\ref{pdcstate}) is not fixed, the condition (\ref{gen_crit}) cannot be directly used to analyze entanglement properties. This can be circumvented by assuming that we have photon number resolving detectors and therefore access to the number of detected photons. This allows us to post-select events that come from the maximally entangled state of an arbitrary number of photons.  After measurement, we have the state
\begin{equation}
|\psi^{(N)}_{\rm max})= \frac{1}{\sqrt{N+1}}  \sum_{m=0}^N |m)_{A} |m)_{B},
\end{equation}
with probability
\begin{equation}
p_N = \frac{1}{\cosh^4 K} \tanh^{2n}K (n+1).
\end{equation}
For the states $|\psi^{(N)}_{\rm max})$ the value of condition (\ref{crit_tmax}) is equal $\epsilon^{(N)} = N+2$ (see Eq. (\ref{statemax})). The average value of $\epsilon$ over different number of emitted photons can be evaluated to be
\begin{eqnarray}
\epsilon_{\rm avg} &=&\sum_{N=0}^{\infty} p_N \epsilon^{(N)} = \frac{1}{16 \cosh^4 K} \\
&\times& (15 \cosh 2K + 6 \cosh 4K + \cosh 6K -6), \nonumber
\end{eqnarray}
which is greater than 1 for all $K>0$. We thus again find that the modified criterion can detect entanglement. We note that the quantity $\epsilon_{\rm avg}$ has only a statistical meaning and cannot be measured in a single run of the experiment.

\section{Gell-Mann versus spin operator basis}
\label{sec:comp}

In the criterion (\ref{gen_crit}) we employed a decomposition to the space spanned by the Gell-Mann matrices (\ref{ctensor}).  One may consider an alternative approach based only on spin operators, which could be equally used to defining an entanglement criterion.  We explain in this section the shortcomings of such an approach and the advantages of the general case (\ref{gen_crit}).

\begin{figure}
\begin{center}
\includegraphics[width=0.48\textwidth]{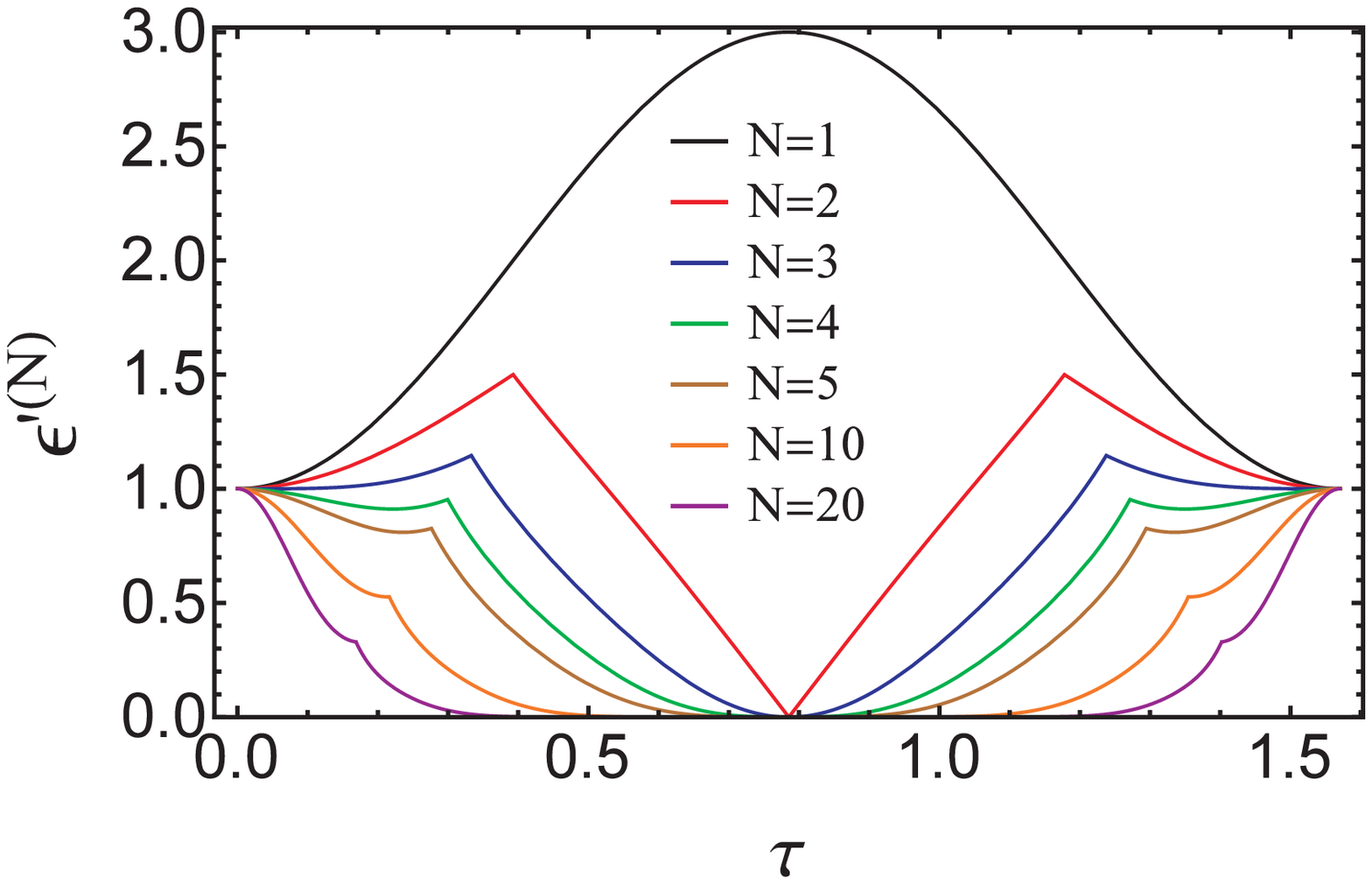}
\caption{\label{fspins} (Color online). The values of $\epsilon^{'(N)}$ parameter for $N\leq 20$. If $\epsilon^{'(N)}>1$, entanglement of the BEC state can be revealed by means of the condition based on the spin operators. }
\end{center}
\end{figure}

Let us define 
\begin{align}
T_{ij}' = \frac{1}{N^2}{\Tr}(\rho S^i \otimes S^j),
\end{align}
where $S^{x,y,z}$ are the Schwinger boson operators (\ref{schwingerboson}).  
The condition for entanglement has the same form as (\ref{gen_crit}),
except that now 
\begin{align}
\mathcal{T}^{'(N)}  = \sum_{i,j=\{x,y,z\}} T_{ij}^{'2}
\label{spinversion}
\end{align}
and $T'_{\rm max}$ refers to the correlation tensor $T'$. While this is a perfectly valid entanglement criterion, we can examine
why such an approach may not be as useful as the general case (\ref{gen_crit}). 
As an example, let us consider the BEC state (\ref{BECstate}) discussed above. Its correlation tensor $T'$ has the following non-vanishing elements 
\begin{eqnarray}
T'_{xx} &=&  \cos^{2N-2}( 2 \tau),\\
T'_{yz} &=& T'_{zy} =  \cos^{N-1}(2 \tau) \sin (2\tau).
\end{eqnarray}
This yields
\begin{align}
\mathcal{T}^{'(N)} = \frac{1}{2}  \cos^{2N-2}{2\tau} (\cos^{2N}{2\tau} +\sin^2{4\tau}).
\end{align}
After diagonalization the correlation tensor $T'$ has the form $\{T'_{xx},T'_{yz},-T'_{zy}\}$,  and
$T'_{\rm max}$ is simply equal to $\max(T'_{xx},|T'_{yz}|)$. We plot the results of using such a criterion in Fig. \ref{fspins} 
for various $ N $.  While for $ N = 1 $ such a criterion works (i.e $\epsilon^{'(N)} > 1$) for all $ \tau $, for $ N = 2, 3 $ the criterion only works in restricted regions of $ \tau $. For  $N > 3$ it completely fails and 
one cannot reveal entanglement for all $\tau$.  This is in contrast to the Gell-Mann approach (Fig. \ref{figT1}) where entanglement can be detected for all $ N $.

The reason why we observe the difference between the approaches utilizing the Gell-Mann matrices and the spin operators is 
that three Schwinger operators ($S_x, S_y$ and $S_z$) do not span the entire space of the dimension $d=N+1$. Moreover, if we express the Schwinger operators in the number state basis, we observe that each component of the spin-$\frac{N}{2}$ operator can be expressed in terms of $N (=d-1)$ Gell-Mann matrices: 
\begin{eqnarray}
S^x &=& \frac{1}{2}\sum_{k=1}^{d-1}   \sqrt{k(d - k)}~M_{\frac{1}{2}(2d(k-1)-k^2+k+2)},\\
S^y &=& \frac{1}{2} \sum_{k=1}^{d-1}  \sqrt{k(d - k)}~ M_{\frac{d(d-1)}{2}+\frac{k(k-1)}{2}+1},\\
S^z &=& \frac{1}{2}\sum_{k=1}^{d-1}  \sqrt{\frac{k(k+1)}{2}} ~M_{d(d-1)+k}.
\end{eqnarray}
For example, for $d=3$
\begin{eqnarray}
S^x &=& \frac{1}{\sqrt{2}}\left(\begin{array}{ccc}0 & 1 & 0  \\ 1 &0 & 1 \\ 0 & 1 &  0\end{array}\right) = \frac{1}{\sqrt{2}}(M_1 + M_3), \nonumber\\
S^y &=& \frac{1}{\sqrt{2}}\left(\begin{array}{ccc} 0 & -i &0 \\ i & 0 & -i \\ 0 & i & 0 \end{array}\right) = \frac{1}{\sqrt{2}}(M_4 + M_5), \label{spin1}\\
S^z &=& \left(\begin{array}{ccc}1 & 0 & 0 \\ 0 & 0 & 0\\0 & 0& -1\end{array}\right) = \frac{1}{2}(M_7 + \sqrt{3} M_8). \nonumber
\end{eqnarray}
This means that in order to describe all three components of the spin operator $S$ we need only $3(d-1)$ out of $d^2-1$ Gell-Mann operators.
However, considering the higher powers of the spin operators one can build a complete operator basis \cite{MENDAS} and  use it to express all Gell-Mann matrices (see Appendix \ref{S2M}). Thus while the criterion based on spin operators can in principle detect entanglement, higher powers of the spin operators would be in general necessary to detect entanglement.  Using only the linear terms of the spin operators appears to miss out in capturing the necessary ingredients to show entanglement is present, at least for the cases that were examined in this study.

\section{Conclusions}
\label{sec:conc}

We constructed a correlation-based criterion for detecting entanglement between two separate (distinguishable) systems of bosons, which are indistinguishable within each subsystem. This is achieved by a change of basis from the standard occupation number basis to the effective $N+1$ dimensional one, which can be treated as the eigenbasis of the $N+1$-dimensional diagonal Gell-Mann matrix. This operation allows for the first time the application of the entire theory of entanglement detection based on correlation tensors. We have applied our theory to several realistic experimental examples, such as $ S^z S^z $ entangled BECs, a light-induced two-mode squeezed BECs, and photons from parametric down conversion.  We find that in each case our criterion effectively detects entanglement for the full range of parameters.  

In order to evaluate the criterion in complete generality, a full set of  Gell-Mann correlations would be required, which would be a challenging task experimentally.  It is possible to derive an analogous criterion (\ref{spinversion}) based only on spin operators which would considerably simplify the detection process.  Unfortunately this criterion appears to be highly insensitive, failing to detect entanglement altogether for large particle systems.  This would suggest that correlations beyond lowest order in the spin operators are necessary to detect entanglement. Using the full set of Gell-Mann correlations gives a much stronger signal of entanglement, and thus this approach appears a more robust method of detecting entanglement. One advantage that our scheme has over other approaches (e.g. using negativity \cite{vidal02})  is that it is based on correlations, rather than tomographically reconstructing the full density matrix.  Thus once a sufficient number of correlations are measured, it gives a direct proof of entanglement, even if the complete set is not measured. Exactly what correlations out of the full Gell-Mann matrix set capture a given type of entanglement is a more difficult task and is an open question.

\acknowledgments

The work is a part of EU project BRISQ2. WL, MM and KK are supported by
NCN Grant No. 2012/05/E/ST2/02352. 
TB thanks G{\'e}za T{\'o}th for discussions. TB and DR are supported by the Transdisciplinary Research Integration Center, Inamori Foundation, Shanghai Research Challenge Fund, and NTT Basic Laboratories. AK acknowledges financial support from the Foundation for Polish Science TEAM project
co-financed by the EU European Regional Development Fund.

\appendix

\section{The Gell-Mann matrices for $d=3$ in terms of components of the spin operator}

\label{S2M}

Let $S^x, S^y$ and $S^z$ be the components of the spin-1 operator presented in Eq. (\ref{spin1}). The set of the three-dimensional Gell-Mann matrices can be expressed in terms of those operators in the following way:
\begin{eqnarray}
M_1 &=& \left(\begin{array}{ccc}0 & 1 & 0 \\ 1 &0 & 0 \\ 0 & 0 &  0\end{array}\right) = \frac{1}{\sqrt{2}} (S^x +S^x S^z +S^z S^x), \nonumber\\
M_2 &=& \left(\begin{array}{ccc}0 & 0 & 1 \\ 0 &0 & 0 \\ 1 & 0 &  0\end{array}\right) = (S^x)^2 - (S^y)^2, \nonumber\\
M_3 &=& \left(\begin{array}{ccc}0 & 0 & 0 \\ 0 &0 & 1 \\ 0 & 1 &  0\end{array}\right) = \frac{1}{\sqrt{2}} (S^x -S^x S^z -S^z S^x), \nonumber\\
M_4 &=&  \left(\begin{array}{ccc}0 & -i & 0 \\ i &0 & 0 \\ 0 & 0 &  0\end{array}\right) = \frac{1}{\sqrt{2}} (S^y +S^y S^z +S^z S^y), \nonumber\\
\end{eqnarray}
\begin{eqnarray}
M_5 &=& \left(\begin{array}{ccc}0 & 0 & 0 \\ 0 &0 & -i \\ 0 & i &  0\end{array}\right) = \frac{1}{\sqrt{2}} (S^y -S^y S^z -S^z S^y), \nonumber\\
M_6 &=&  \left(\begin{array}{ccc}0 & 0 & -i \\ 0 &0 & 0 \\ i & 0 &  0\end{array}\right) = S^x S^y +S^y S^x, \nonumber\\
M_7 &=& \left(\begin{array}{ccc}1 & 0 & 0 \\ 0 &-1 & 0 \\ 0 & 0 &  0\end{array}\right) = \frac{1}{2} (S^z +2 (S^z)^2 - (S^x)^2 - (S^y)^2), \nonumber\\
M_8 &=& \frac{1}{\sqrt{3}} \left(\begin{array}{ccc}1 & 0 & 0 \\ 0 &1 & 0 \\ 0 & 0 & -2\end{array}\right) = \frac{1}{2\sqrt{3}} (3 S^z +2(S^z)^2 + (S^x)^2 + (S^y)^2). \nonumber
\end{eqnarray}


\begin{thebibliography}{99}
\bibitem{footnote1}
The highest Schmidt coefficient, $T_{\rm max}$, can be calculated as the square root of the  highest eigenvalue of $TT^\dagger$.
\end{thebibliography}
\end{document}